# Transceiving signals by mechanical resonance: A low frequency (LF) magnetoelectric mechanical antenna pair with integrated DC magnetic bias

Yunping Niu[1,2,3] and Hao Ren[1,*]

[1] School of Information Science and Technology, ShanghaiTech University, Shanghai, 201210, China

[2] Shanghai Institute of Microsystem and Information Technology, Chinese Academy of Sciences, Shanghai, 200050, China.

[3] University of Chinese Academy of Sciences, Beijing, 100049, China.

*Corresponding author E-mail: renhao@shanghaitech.edu.cn

**Abstract:**

Low frequency communication systems offer significant potential in portable electronics and internet of things (IoT) applications due to the low propagation loss and long transmission range. However, because the dimension of electrical antenna is comparable to one quarter wavelength of the electromagnetic wave that it transmits and receives, currently a gap exists between LF communication systems and IoT, as IoT generally implement miniaturized electrical antennas with a small wavelength at a high frequency while traditional LF electrical antennas are too bulky. In this paper, we present an LF magnetoelectric mechanical transmitting and receiving antenna pair to significantly miniaturize the dimension of LF antennas. As the operation principle of the magnetoelectric mechanical antenna pair is based on mechanical resonance, its dimension is reduced by four orders of magnitude compared with an electrical antenna counterpart. The transmitting and receiving antennas are the same in structure and dimensions: they are both composed of magnetostrictive Terfenol-D and piezoelectric PZT laminate, and their dimensions are both $38 \times 12 \times 8.2 mm^3$. Both the transmitting and receiving antennas are integrated with DC magnetic bias to improve its performance. Antenna pair performance measurement demonstrates that a maximum operation distance of 9m is demonstrated with DC magnetic bias, while maximum operation distance without DC magnetic bias is 4m, which is reduced by 56%. The LF magnetoelectric mechanical antenna pair may be a promising candidate for the portable electronics and IoT applications.

Radio-frequency communication has been a critical component for human society since its invention in the late nineteenth century[1, 2]. A variety of radio-frequency communication systems, such as medium frequency(MF) for AM radio communications, ultra-high frequency(UHF) communication systems for television broadcasting and cellphone communications[3-7], and mmWave for 5G communications and mmWave radar[8-10] have been widely implemented based on scientific and technological advancements, which greatly improve our daily life. As the interface between radio waves propagating through space and electric currents moving in metal conductors, antennas have emerged since the beginning of 20$^{th}$ century as one of the most critical components in radio-frequency communications. The earliest antennas were wire type, primarily as single elements[11]. In the past 40 years, advanced antennas including microstrip antennas, dielectric resonator antennas, and multiple input-multiple output (MIMO) technology are extensively studied[12-14]. The microstrip antennas are applied in television broadcasting[15] and 5G WLAN application[16]. The waveguide slot antennas and MIMO antennas are applied in 5G communication systems and mmWave radar[17-19].

Low frequency(LF) antennas, which operate at frequencies between 30-300kHz, have been widely implemented in meteorological broadcasts, radio navigation, and underwater communication[20, 21]. Compared with antennas operating at higher frequency bands, the LF antennas are not sensitive to dielectric loss in water due to their long wavelength(1-10km). As a result, LF antennas are suitable for underwater communications. Due to the need of connecting to various sensors, actuators and microsystems to collect and send data[22, 23], internet of things(IoT) have been a research focus and it requires miniaturized antennas for wireless communication system. Nowadays antenna miniaturization remains as one of the major critical issues for IoT research. The size of traditional electrical antenna is comparable to the wavelength($\lambda$), which normally varies from 0.1 to 0.5$\lambda$ to transmit the electromagnetic wave effectively[24], and a miniaturized antenna operates at a high frequency, which causes the electromagnetic wave to attenuate significantly during its transmission. On the other hand, the large dimension of LF antennas makes them difficult to be utilized in portable electronics, wireless sensor networks and IoT[25]. As a result, one of the key challenges on state-of-art LF antennas lies in their miniaturization[26].

In the past few years, magnetoelectric(ME) antennas are proposed as an alternative to electrical antenna to miniaturize the antenna dimension[26-32]. The ME effect is an electric polarization response to an applied magnetic field, or conversely a magnetization response to an applied electric field[33, 34], and it has found applications in gyrators and energy converters[35-40]. These ME antennas transmit and receive electromagnetic waves through the magnetoelectric effect, which converts electric power into mechanical resonance, then converts mechanical resonance into alternating magnetic field or vice versa. The ME antennas have small dimensions and can be utilized in portable electronics and IoT. As the ME coefficient is related to the bias DC magnetic field: it rises first then falls and reaches a maximum under an optimal bias magnetic field, prior studies utilized electromagnets or Helmholtz coils to provide a magnetic bias to improve the performance of ME devices[41-44]. However, the electromagnets and Helmholtz coils are too bulky to miniaturize the system. In order to enhance the performance with a small footprint, Niu and Ren integrated a DC magnetic field bias to an ME receiving antenna[45]. However, no prior art has been reported to integrate a miniaturized DC magnetic bias to significantly improve the performance of a ME transmitting antenna.

In this work, we present a pair of miniaturized ME transmitting (Tx) and receiving (Rx) LF ME mechanical antenna operating at a mechanical resonant frequency of 37.95kHz. In order to improve the performance with a small dimension, we integrate miniaturized integrated Rb magnets to provide a bias DC magnetic field and demonstrate that the integrated Rb magnets can improve the performance of Tx-Rx ME antenna pair significantly.

Figure 1(a) and (b) shows the schematic and the optical image of the ME mechanical antenna pair. Figure 1(c) depicts a typical testing platform for the ME mechanical antenna pair. The ME antenna has a sandwich structure of a PZT layer between two TbDyFe$_2$(Terfenol-D) layers. The PZT layer is a stack of 30 pieces of 0.2mm-thick PZT laminates and the total thickness is 7mm. The Terfenol-D layer is glued at both sides of PZT layer by epoxy adhesives. The dimensions of PZT and Terfenol-D layers are $38\times12\times7mm^3$ and $35\times8\times1.2mm^3$. Four additional devices with different PZT thickness (1-layer PZT of 0.25mm, 3-layer PZT of 0.8mm, 7-layer PZT of 1.5mm and 15-layer PZT of 3.4mm) are fabricated. Two wires are bonded on both sides of the PZT layers as interconnects. Four $10\times10\times10mm^3$ Rb magnets are placed on the Tx and Rx antennas to improve their performance. The schematic of a test platform for typical application scenarios is illustrated in Figure 1(b). A sine wave signal is generated by a waveform generator (DG1062Z, RIGOL

Technology Inc.) and amplified by a high voltage power amplifier (ATA-2041A, Aigtek Inc.). The amplified signal is fed to the transmitting antenna to generate AC electromagnetic wave. The electromagnetic wave is received by the receiving antenna and then amplified by a low-noise preamplifier (SR560, Stanford Research Systems Inc.). The amplified signal is fed to a RF spectrum analyzer (FSW8, Rohde & Schwarz GmbH.) or oscilloscope (MSO2024B, Tektronics Inc.) for signal recording.

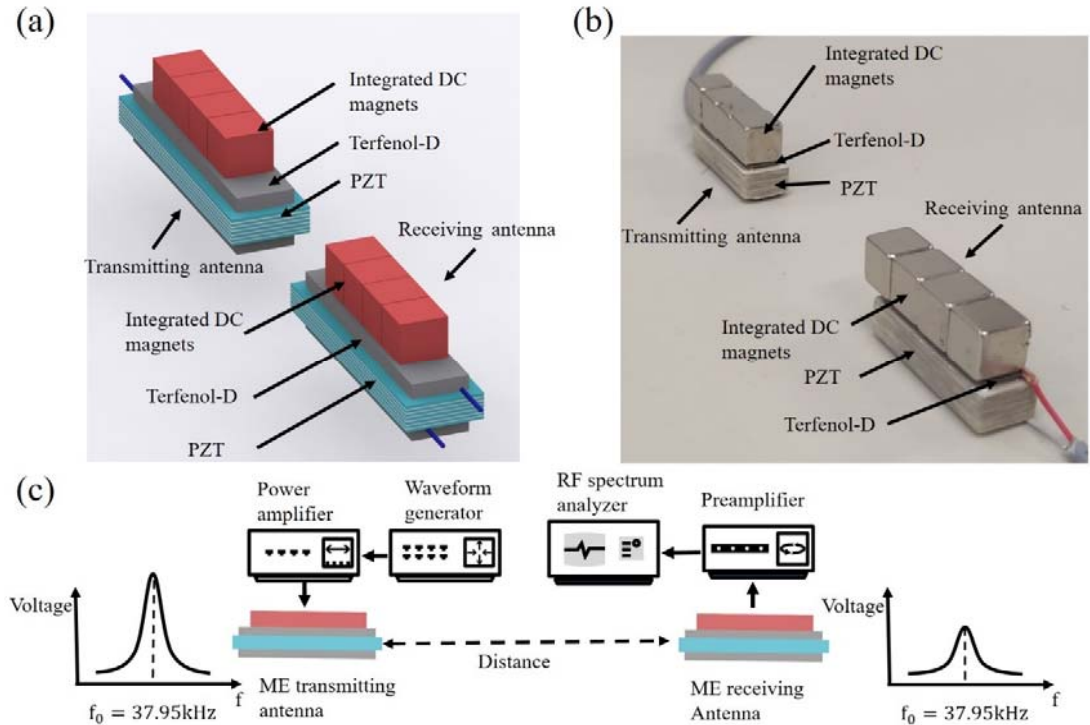

**Figure 1 (a) The schematic of the ME transmitting/receiving antenna pair; (b) The optical image of the ME transmitting/receiving antenna pair; (c) The schematic of a typical testing platform for the ME antenna pair.**

The ME antenna is based on the ME effect[33, 34]. Magnetostrictive and piezoelectric laminated composites show significant ME effect[33, 42]. In this paper, Terfenol-D and PZT are chosen as magnetostrictive and piezoelectric material, respectively. PZT is a piezoelectric material, which generates a voltage when a stress is applied, or in inverse piezoelectric effect, a stress is generated in the piezoelectric material when a voltage is applied. Terfenol-D is a material with a giant magnetostrictive effect[42, 44, 46, 47], and it deforms under a magnetic field. For the inverse magnetostrictive effect, a magnetic field is generated when a stress is applied a magnetostrictive material. We propose the operation principle of the ME transmitter/receiver pair to be piezoelectromagnetic effect, as applying an AC voltage on the transmitter generates an alternating strain and an electromagnetic wave while applying an electromagnetic wave on the receiver generates an alternating strain and a voltage. When the input transmitting electrical signal at the resonant frequency of the antenna is applied to both sides of PZT layers of the Tx antenna, an alternating strain is generated in the PZT because of the piezoelectric effect. As the PZT layer and Terfenol-D layer are bonded together as laminate, the Terfenol-D layers senses the strain. As a result, the ME antenna produces a mechanical resonance at the same frequency of the transmitting signal. Due to the magnetization oscillation of the magnetostrictive layer, an alternating electromagnetic wave is generated. In the meantime, the mechanical resonance of the PZT layer results in electrical dipole oscillation, which also generates an alternating electromagnetic wave. At the Rx antenna, the process is reversed. The Terfenol-D layer extends and contracts under the electromagnetic wave. The strain drives the PZT layer to generate an alternating voltage signal between its two electrodes. In this way the signal is transmitted at the Tx antenna and is received at the Rx antenna.

To achieve a maximum output voltage and operation distance, the frequency of the transmitting signal should be the same as the intrinsic longitudinal mechanical resonant frequency of the ME mechanical antenna. The i-th resonant frequency in the longitudinal direction of a multi-layer laminated rectangular plate is calculated by the following equation:

$$f_i = (2i-1)\frac{1}{2L}\sqrt{\frac{CA_e}{\rho A_e}}$$
$$CA_e = C_1 A_1 + C_2 A_2$$
$$\rho A_e = \rho_1 A_1 + \rho_2 A_2$$
(1)

where $L$ is the length of the structure. $A$ is the cross-sectional area, $CA_e$ and $\rho A_e$ represent the effective Young's module and equivalent mass of the structure. The Young's module of Terfenol-D and PZT are 50GPa and 82.1GPa, the density of Terfenol-D and PZT are 9250 kg/m$^3$ and 7750 kg/m$^3$. The calculated resonant frequency is 39.66 kHz, in agreement with the experimental result by measuring the resonant frequency of the ME antenna by an impedance analyzer (E4990A, Keysight Technology Inc.) in Figure 2.

Figure 2(a) shows the impedance and phase of the ME antenna. The ME antenna resonates at 37.95kHz, which is slightly lower than the analytical resonant frequency based on equation 1. The frequency response of the ME antenna is shown in Figure 2(b). When the operation frequency is shifted to 35.8kHz and 38.65kHz, the normalized output voltage decreases to 0.707 of maximum voltage. The bandwidth of ME antenna is narrow since when the operating frequency deviates from the resonant frequency, the strain of the Tx antenna decreases and the transmitted electromagnetic field decreases. Meanwhile, the strain of the receiving antenna decreases as well. The decrease in both Tx and Rx occurs when the operating frequency deviates from the resonant frequency, and the output voltage is reduced significantly.

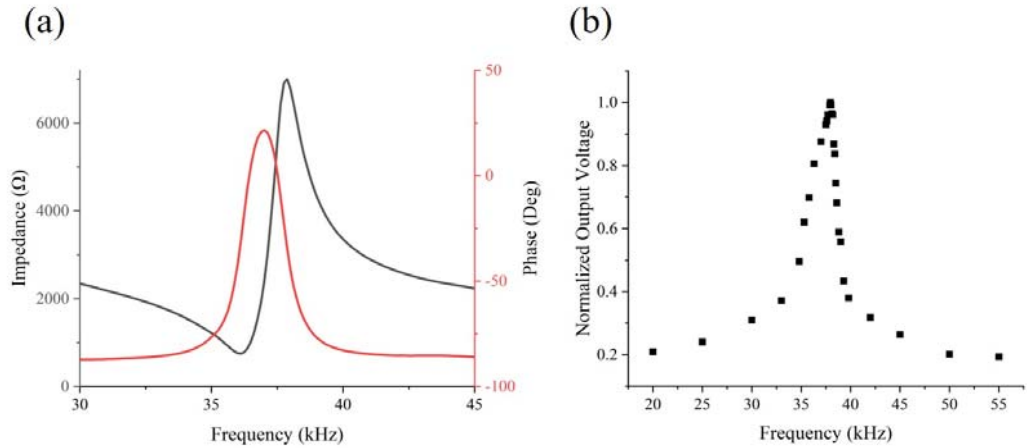

**Figure 2 (a) The impedance versus frequency of ME antenna; (b) The normalized output voltage as a function of input signal frequency.**

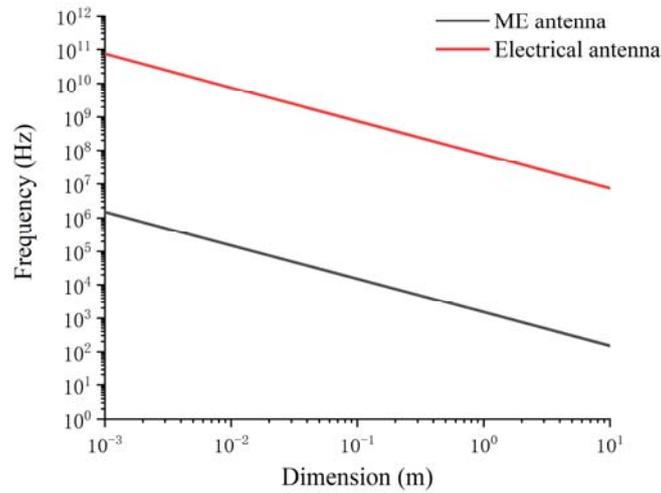

**Figure 3 The comparison of the resonant frequencies of the ME antenna and the electrical antenna.**

Figure 3 shows the comparison of the resonant frequencies of the ME antenna and an electrical antenna. The relationship between resonant frequency and dimension of ME antenna follows Equation 1. The dimension of the electrical antenna is a quarter of wavelength and the frequency is calculated by $f = c/\lambda$. Here $f$, $c$ and $\lambda$ are frequency, speed of light, and wavelength, respectively. According to Figure 3, the dimension of the ME antenna is more than four orders of magnitude smaller than the electrical antenna.

To enhance the ME effect and the performance, a DC magnetic field bias is applied to the ME antenna. Previous studies demonstrated the ME effect varies with DC bias magnetic field and reaches a maximum under an optimal bias magnetic field[33, 34, 42, 44, 48]. Therefore, many studies adopted bulky electromagnets or Helmholtz coils to enhance the output voltage and efficiency of ME gyrators or energy converters. However, the electromagnets or Helmholtz coils have large dimensions. To minimize the size of the DC magnetic bias, we adopt four Rb magnets on the Tx and Rx antenna to provide DC magnetic field bias. To verify the effect of Rb magnets, the antenna is positioned in an electromagnet along its longitudinal direction. As illustrated in the output voltage verse magnetic field plot in Figure 4, the output voltage first rises then falls with an increasing magnetic field in both forward and reverse direction. When the antenna is not integrated with magnets, the output voltage rises between 0~660Oe and falls from 833Oe, reaching a maximum between 660~833Oe. The trend is similar in the reverse direction: the maximum voltage occurs between -833~-660Oe. The magnetic field of the Rb magnets is 320Oe measured by a gauss meter. When the antenna is integrated with the magnets, the curve shifts left by 320Oe because the magnets provide a magnetic field of 320Oe. In this case, the output voltage increases at zero DC magnetic bias. The experiment demonstrates that the Rb magnets improve the performance of the ME antenna. Compared with traditional electromagnets or Helmholtz coils, the Rb magnets significantly reduce the dimension.

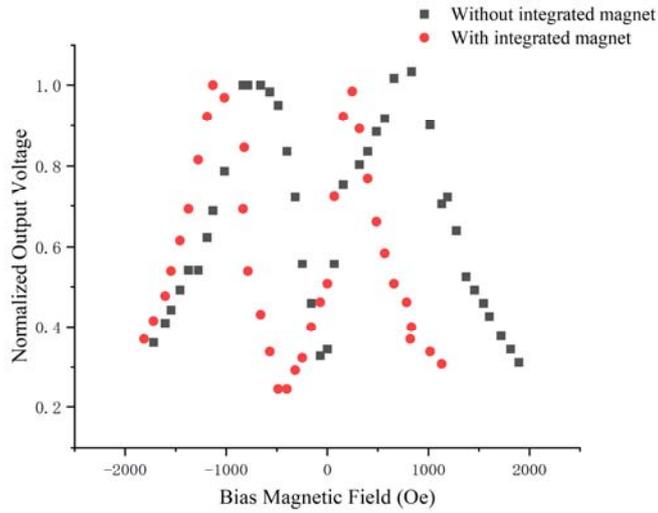

**Figure 4 Normalized output voltage as a function of magnetic field bias with and without Rb magnets.**

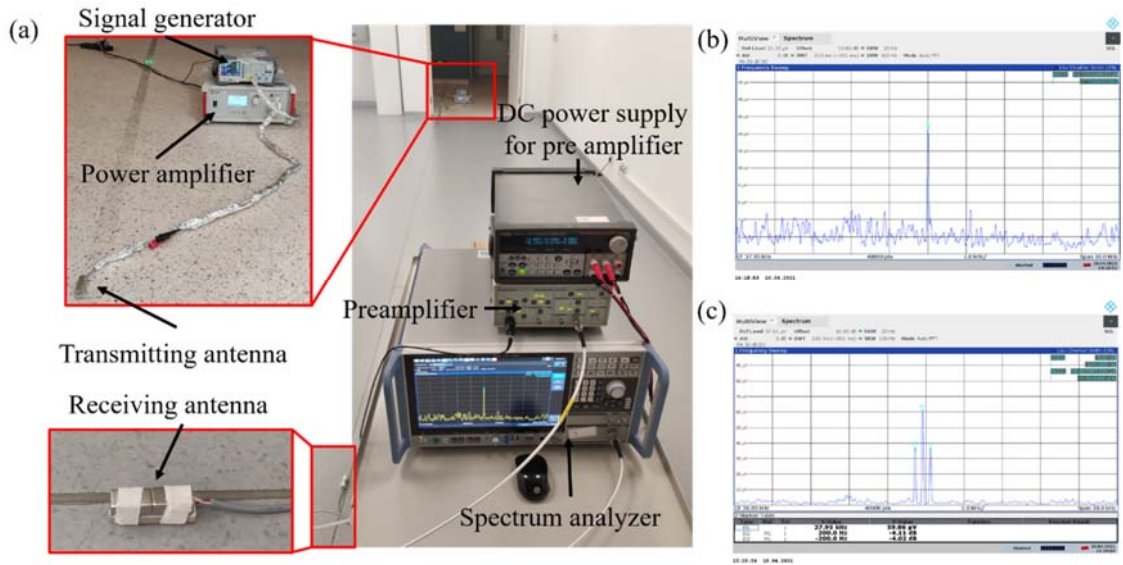

**Figure 5 (a)Optical image of the ME antenna pair measurement setup; (b) RF spectrum of the received signal from the ME antenna at 9m; (c) RF spectrum of the received signal from the ME antenna at 3m with an 200Hz sine wave AM modulation.**

The optical image of the ME antenna pair measurement setup for a typical application scenario is shown in Figure 5(a). A 37.95 kHz sine wave is generated by the waveform generator and fed to the high voltage amplifier. The amplified signal has an amplitude of 400Vpp. At 37.95 kHz, the impedance of the transmitting antenna is $Z= 7000e^{-j44°}\Omega$, as shown in Figure 2(a). The input power is calculated as $P_{in} = \frac{V_{rms}^2}{|Z|} \cos\theta =$ 2.03W. The RF spectrum of the receiving antenna is shown in Figure 5(b) and Figure 5(c). When the distance between the Tx and Rx antennas is 9m, the spectrum is shown in Figure 5(b), which shows that the signal to noise ratio is 13.6dB, with received signal power level of -83.8dBm and the noise level of -97.4dBm. Figure 5(c) shows the spectrum of amplitude modulated signal at 3m. The frequency of carrier wave is 37.95kHz to obtain the maximum voltage. The modulation signal is 200Hz sine wave and the modulation depth is 100%.

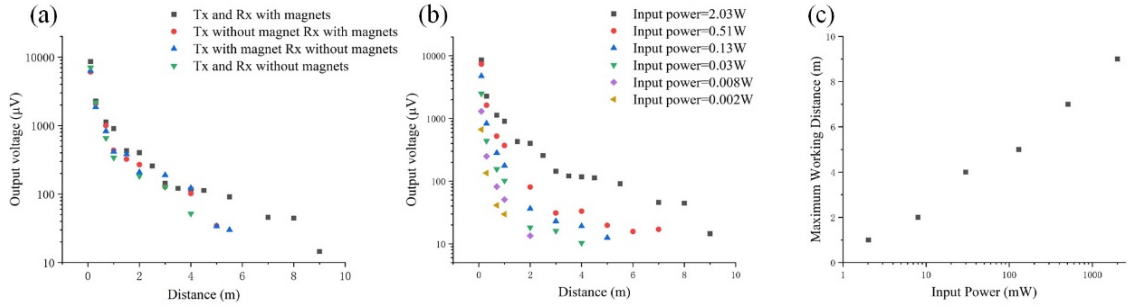

**Figure 6 (a) Output voltage as a function of distance between Tx and Rx antenna with and without magnets; (b) Output voltage as a function of distance between Tx and Rx antenna under different input power; (c) Maximum operation distance as a function of input power of Tx antenna.**

Figure 6(a) shows the output voltage of the Rx antenna at different distances between the Tx and Rx antennas and the influence of the integrated magnets on the output voltage. The voltage drops significantly with the increase of distance. At a distance of 1m, the output voltage is 903μV for the Tx and Rx antennas with integrated magnets. When the distance increases to 9m, the output voltage decreases to 14.51μV, which is only 1.6% of the output voltage at 1m. The output voltage also decreases as the magnets are removed. At a distance of 1m, the output voltage when neither the Tx nor the Rx antenna is integrated with magnets is 340μV, which is reduced by 62.3%. The output voltage when the Rx antenna is integrated with magnets while the Tx antenna is not integrated with magnets is 435μV, which is reduced by 51.8% compared with the output voltage when both Tx and Rx antennas are integrated with magnets. The output voltage when the Tx antenna is integrated with magnets while the Rx antenna is not integrated with magnets is 415.9μV, which is reduced by 53.9% compared with the output voltage when both the Tx and Rx antennas are integrated with magnets. Not only does the amplitude of the output voltage decrease, but the maximum operation distance also decreases. The maximum operation distance is 9m when the Tx and Rx antenna are both integrated with magnets. When neither the Tx nor the Rx antenna is integrated with magnets, the maximum detection distance is 4m, which is reduced by 56%. This result demonstrates that the integrated magnets improve the performance of the ME antenna, which is in agreement with the result of Figure 4. Figure 6(b) shows the output voltage of Rx antenna at different distances between Tx and Rx antenna at different input power. Figure 6(b) and Figure 6(c) demonstrates that with higher input power, the output voltage and maximum operation distance increases significantly. When the input power is 0.002W and 2W, the output voltage at 1m is 29.7μV and 903μV respectively and the maximum operation distance is 1m and 9m respectively, which is increased by 30 times and 9 times.

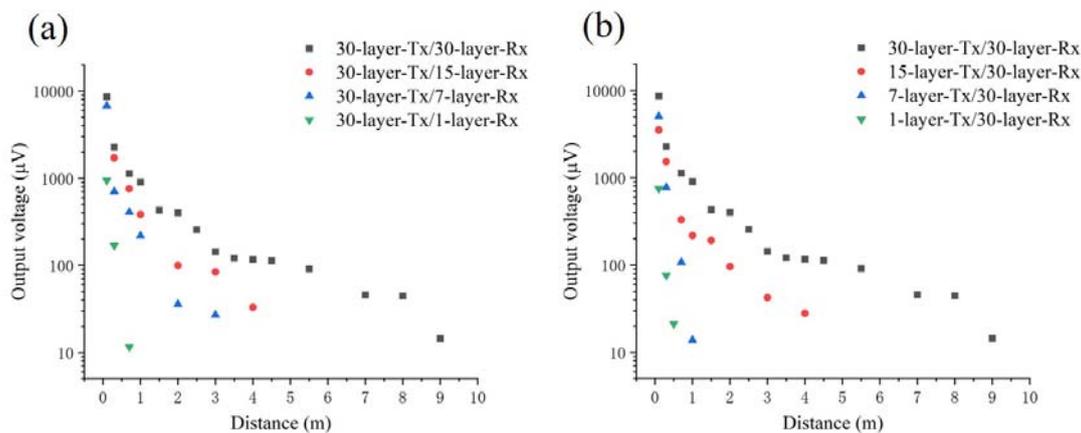

**Figure 7 Output voltage varies with distance under different layers of the Tx and Rx antenna.**

Figure 7 shows how the output voltage varies with the distance when the number of PZT layers of the Tx and Rx antennas changes. The experiment results demonstrate that for both the Tx and the Rx antenna, when the number of PZT layer decreases, the output voltage and the maximum operation distance decreases significantly. At 0.1m distance, the output voltage of 30-layer-Tx/30-layer-Rx antenna is 8.5mV, while the output voltage of 30-layer-Tx/1-layer-Rx antenna and 1-layer-Tx/30-layer-Rx antenna is 0.943mV and 0.751mV, respectively, which is reduced by 89% and 91%. The reason for the increase in voltage when the number of PZT layers increases is all these PZT laminates are equivalent to capacitors, when they are stacked in series, the voltage adds up to increase the output voltage.

We also characterize the efficiency and quality factor of the ME antenna. Due to the low operation frequency and the long wavelength, it is difficult to directly measure the transmitted power to calculate the efficiency of the ME antennas. Thus the efficiency is measured by comparing with a theoretical circular loop antenna[32]. The near field flux density of a small circular loop antenna can be derived as[32]

$$|B_r| \approx \frac{\mu_o}{\sqrt{2}} \sqrt{\frac{3}{\pi \eta k^4} \frac{Z_o R_r P_{in}}{(R_r + Z_o)^2}} \left(\frac{2\cos\theta}{r^3}\right)$$

$$|B_\theta| \approx \frac{\mu_o}{\sqrt{2}} \sqrt{\frac{3}{\pi \eta k^4} \frac{Z_o R_r P_{in}}{(R_r + Z_o)^2}} \left(\frac{\sin\theta}{r^3}\right)$$

(2)

where $\mu_0$ is the permeability of the free space($4\pi \times 10^{-7}$H/m), $\eta$ is the free space radiation impedance (377Ω). $k$ is the propagation constant for a wavelength $\lambda$ ($k = \frac{2\pi}{\lambda}$). $Z_0$ is the impedance of the transmission line ($Z_0 = 50\Omega$). $S$ is the area of the circular loop, which is the same as the ME antenna (456mm²). $R_r$ is the radiation resistance of the antenna ($R_r = \frac{8}{3}\eta\pi^3(\frac{S}{\lambda^2})^2 \approx 31171\frac{S^2}{\lambda^4}$). $P_{in}$ is the input power which is the power provided to the ME antenna ($P_{in} = 2.03$W). $r$ is the distance between the observation point and the center of the antenna. $\theta$ is the angle of the observation point as shown in Figure 8, which shows the measured magnetic flux density of the ME antenna and the analytical magnetic flux density of the small circular loop antenna as a function of distance $r$ in two directions, i.e., $\theta = 90°$ and $\theta = 0°$.

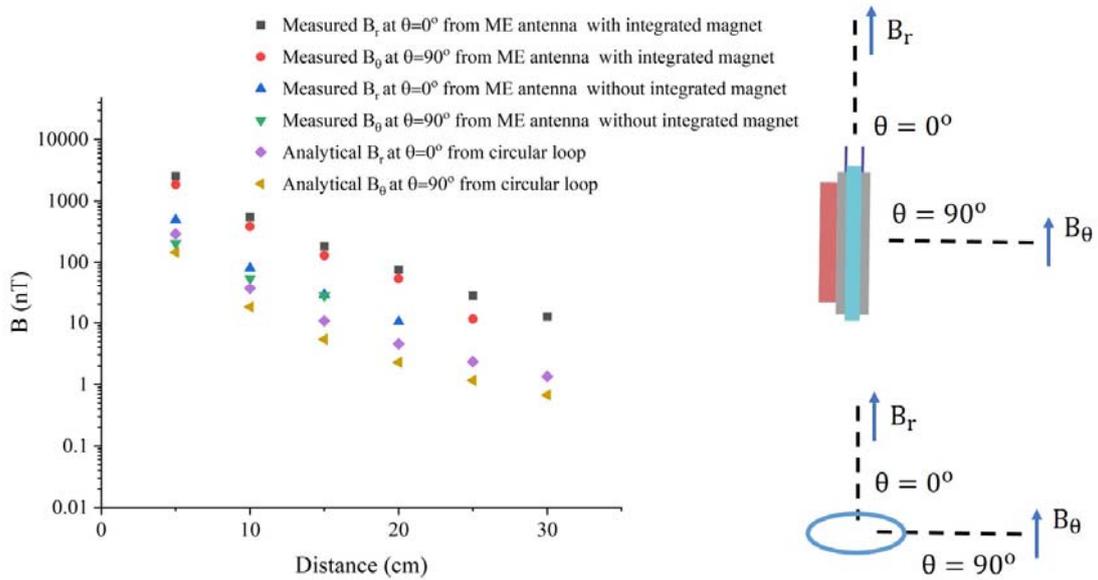

**Figure 8 The measured and analytical magnetic flux density of ME antenna and small circular loop antenna.**

Since the radiation efficiency of an antenna is proportional to the squared magnetic flux density[32], the radiation efficiency of the ME antenna can be derived by comparing the magnetic flux density in Figure 8.

The magnetic flux density of the ME antenna is 23 times larger than the circular loop, therefore the efficiency ratio is.

$$\frac{\eta_{ME}}{\eta_{loop}} = (\frac{|B_{ME}|}{|B_{loop}|})^2 = 529 \qquad (3)$$

The radiation efficiency of a circular antenna can be calculated as follow[32, 49]:

$$\eta_{loop} = \frac{R_r}{R_r + R} \approx \frac{R_r}{R} = 2.6 \times 10^{-20} \qquad (4)$$

The efficiency of the circular loop is extremely low because its size is much smaller than its wavelength and the antenna is non-resonant[50]. The ME antenna operating at mechanical resonance enhances the efficiency by 529 times to $1.37 \times 10^{-17}$. The magnetic flux density of ME antenna with integrated magnets is 8 times larger than ME antenna without integrated magnets, therefore the efficiency ratio is increased by 64 times. The efficiency of the ME antenna is in agreement with previously published work, which is in the order of $10^{-16}$ [32].

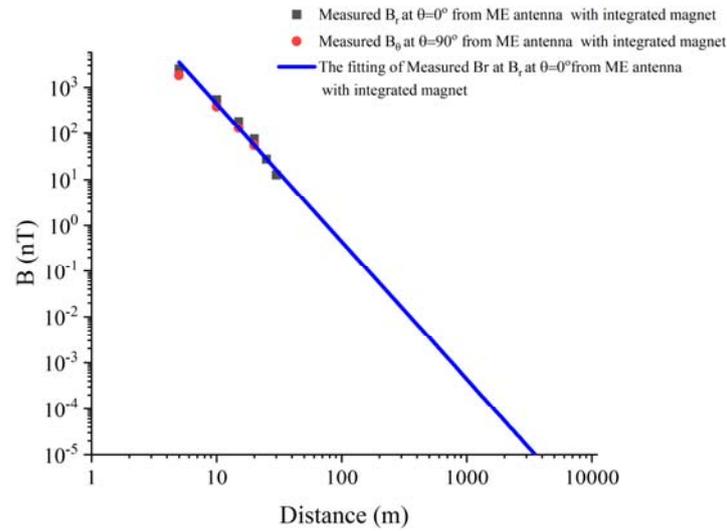

**Figure 9 The measured and fitting magnetic flux density of ME antenna as a function of the distance between the ME transmitter and the receiver.**

Figure 9 shows the measured magnetic flux density of ME antenna and its fitting curve. The fitting curve demonstrates that the measured magnetic flux density is proportional to $1/r^3$, where $r$ is the distance between the Tx and Rx antennas. This trend agrees with the theoretical result in Equation 2.

The quality factor of ME antenna can be calculated as follows:

$$Q_{ME} = \frac{f}{BW_{3-dB}} \qquad (5)$$

As a result, the calculated $Q_{ME}$ is 13.3, which is in agreement with previously published work of 200.2 and 25.4[45, 51].

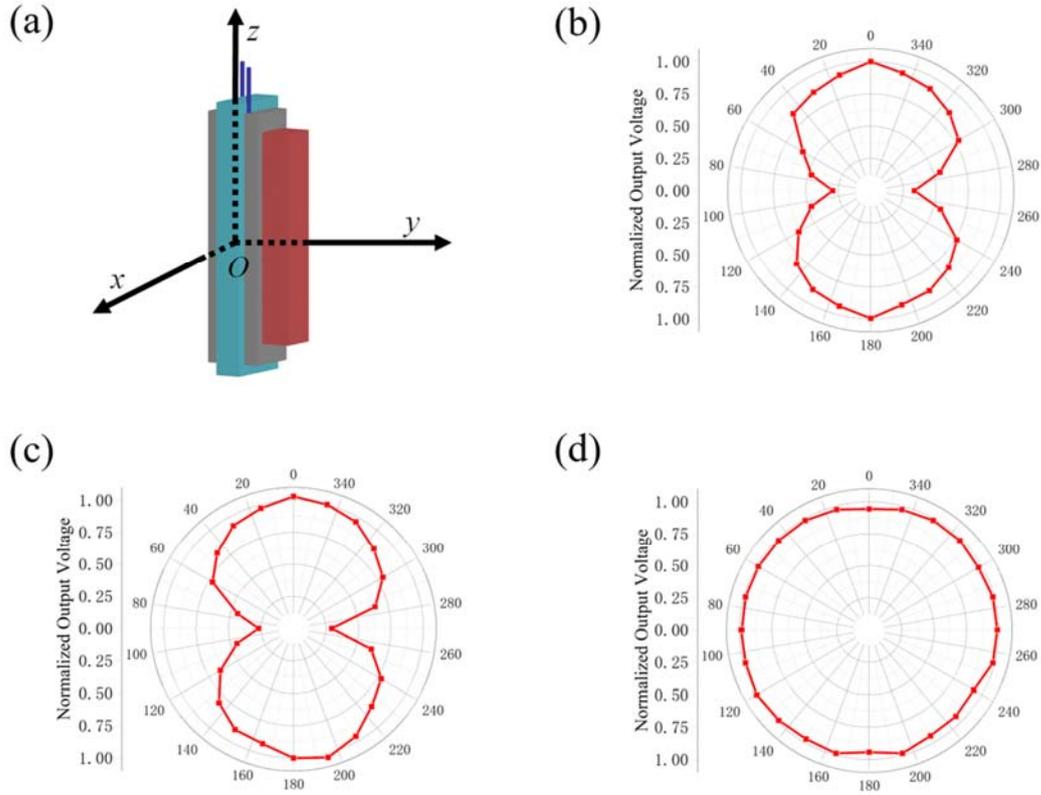

**Figure 10 (a) The schematic diagram of coordinate system and the measured normalized received voltage as a function of different angle around the transmitting antenna in (b) *xOz*-plane; (c)*yOz*-plane; (d) *xOy*-plane.**

To measure the radiation pattern of the ME antenna, we changed the angle of the ME antenna to obtain the normalized output voltage. The coordinate system is set as shown in Figure 10(a): the longitudinal direction of the ME antenna is along the *z* axis, the width direction is along the *x* axis and the thickness direction is along the *y* axis. The output voltage of the antenna is recorded at different angles and the radiation pattern of the three different plane is shown in Figure 10(b), (c) and (d) respectively. The radiation pattern of *xOz* and *yOz* plane is similar to a horizontal small-loop antenna or magnetic-dipole antenna[32]: the maximum occurs at 0 and 180° and the minimum occurs at 90° and 270° and the curve follows the cosine profile. However, at 90° and 270° the voltage is not zero, which may be because our ME antenna cannot be equivalent to an ideal magnetic-dipole antenna and the reflection and interference of the measurement setup. According to Figure 10, the directivity can be estimated as 2.22 (3.46dB). The gain of this antenna can be calculated as $G = \eta_{ME} D = 3 \times 10^{-17}$.

In summary, an LF magnetoelectric mechanical transmitting and receiving antenna pair with an integrated DC magnetic bias is presented in this paper. Both the transmitting and the receiving antenna operates by mechanical resonance to transmit and receive electromagnetic wave. Compared with traditional electrical antennas operating at the same frequency, the dimension of the proposed magnetoelectric antennas is reduced by more than four orders of magnitudes, which makes them to be promising candidates for portable electronics and IoT applications. The integrated DC magnets provide a magnetic bias of 320Oe, which improves the efficiency of the antenna, and the maximum operation distance with magnetic bias is improved by 2.25 times, from 4m to 9m. The efficiency, gain and directivity of the antennas are also characterized. While this paper presents a LF magnetoelectric mechanical antenna pair, magnetoelectric antennas operating at other frequencies can be achieved by adjusting the dimensions of the piezoelectric and magnetostrictive materials by equation 1. The small dimension and good performance of the LF mechanical antenna pair offers great potential for IoT applications.

## Author contribution
H.R. conceived the idea. H.R. and Y.N. designed the experiments. Y.N. implemented the experimental set-up and conducted experiments. Y.N. and H.R. analyzed data and participated in discussions. Y.N. and H.R. wrote the paper. H.R. led the work.

## Reference

1. N. Tesla, Apparatus for transmission of electrical energy, 5/15/1900.
2. T. A. Edison, Improvement in phonograph or speaking machines, 2/19/1878.
3. R. F. J. Broas, D. F. Sievenpiper and E. Yablonovitch, IEEE Transactions on Microwave Theory and Techniques **49** (7), 1262-1265 (2001).
4. W. F. Schreiber, Proceedings of the IEEE **83** (6), 958-981 (1995).
5. D. A. Anderson, R. E. Sapiro and G. Raithel, IEEE Transactions on Antennas and Propagation **69** (5), 2455-2462 (2021).
6. Y. Shimizu, K. Shigeta, K. Yukawa, T. Nakamura, M. Mikkaichi, Y. Nagasawa and R. Sato, IEEE Transactions on Antennas and Propagation **36** (7), 927-935 (1988).
7. J. R. Mensch and C. C. Pearson, IEEE Transactions on Communications Systems **12** (1), 124-125 (1964).
8. T. S. Rappaport, S. Sun, R. Mayzus, H. Zhao, Y. Azar, K. Wang, G. N. Wong, J. K. Schulz, M. Samimi and F. Gutierrez, IEEE Access **1**, 335-349 (2013).
9. Y. Niu, Y. Li, D. Jin, L. Su and A. V. Vasilakos, Wireless Networks **21** (8), 2657-2676 (2015).
10. P. Kumari, J. Choi, N. González-Prelcic and R. W. Heath, IEEE Transactions on Vehicular Technology **67** (4), 3012-3027 (2018).
11. C. A. Balanis, in *2012 IEEE International Workshop on Antenna Technology (iWAT)* (2012), pp. 5-7.
12. R. Munson, IEEE Transactions on Antennas and Propagation **22** (1), 74-78 (1974).
13. S. A. Long and E. M. O. Connor, in *2016 IEEE International Symposium on Antennas and Propagation (APSURSI)* (2016), pp. 679-680.
14. M. A. Jensen, in *2016 IEEE International Symposium on Antennas and Propagation (APSURSI)* (2016), pp. 681-682.
15. A. V. Suraperwata, Hendrawan, J. Suryana and S. Haryadi, in *2014 8th International Conference on Telecommunication Systems Services and Applications (TSSA)* (2014), pp. 1-4.
16. A. Madankar, V. Chakole and S. Khade, in *2020 3rd International Conference on Intelligent Sustainable Systems (ICISS)* (2020), pp. 1142-1144.
17. Z. Ahmed, P. McEvoy and M. J. Ammann, in *2018 IEEE MTT-S International Microwave Workshop Series on 5G Hardware and System Technologies (IMWS-5G)* (2018), pp. 1-3.
18. M. Awais, H. S. Khaliq and W. T. Khan, in *2017 Progress in Electromagnetics Research Symposium - Fall (PIERS - FALL)* (2017), pp. 2802-2807.
19. Z. U. Khan, Q. H. Abbasi, A. Belenguer, T. H. Loh and A. Alomainy, in *2018 IEEE MTT-S International Microwave Workshop Series on 5G Hardware and System Technologies (IMWS-5G)* (2018), pp. 1-3.
20. R. L. Frank, Proceedings of the IEEE **71** (10), 1127-1139 (1983).
21. P. Lunkenheimer, S. Emmert, R. Gulich, M. Köhler, M. Wolf, M. Schwab and A. Loidl, Physical Review E **96** (6), 062607 (2017).
22. Y. Koga and M. Kai, in *2018 IEEE-APS Topical Conference on Antennas and Propagation in Wireless Communications (APWC)* (2018), pp. 762-765.
23. G. Bedi, G. K. Venayagamoorthy, R. Singh, R. R. Brooks and K. Wang, IEEE Internet of Things Journal **5** (2), 847-870 (2018).
24. D. B. Miron, in *Small Antenna Design*, edited by D. B. Miron (Newnes, Burlington, 2006), pp. 1-8.
25. D. E. Hurdsman, P. M. Hansen and J. W. Rockway, in *IEEE Antennas and Propagation Society International Symposium. Digest. Held in conjunction with: USNC/CNC/URSI North American Radio Sci. Meeting (Cat. No.03CH37450)* (2003), Vol. 4, pp. 811-814 vol.814.

26. T. Nan, H. Lin, Y. Gao, A. Matyushov, G. Yu, H. Chen, N. Sun, S. Wei, Z. Wang, M. Li, X. Wang, A. Belkessam, R. Guo, B. Chen, J. Zhou, Z. Qian, Y. Hui, M. Rinaldi, M. E. McConney, B. M. Howe, Z. Hu, J. G. Jones, G. J. Brown and N. X. Sun, Nat Commun **8** (1), 296 (2017).
27. C. Dong, Y. He, M. Li, C. Tu, Z. Chu, X. Liang, H. Chen, Y. Wei, M. Zaeimbashi, X. Wang, H. Lin, Y. Gao and N. X. Sun, IEEE Antennas and Wireless Propagation Letters **19** (3), 398-402 (2020).
28. H. Lin, M. Zaeimbashi, N. Sun, X. Liang, H. Chen, C. Dong, A. Matyushov, X. Wang, Y. Guo, Y. Gao and N. Sun, in *2018 IEEE/MTT-S International Microwave Symposium - IMS* (2018), pp. 220-223.
29. Z. Yao, Y. E. Wang, S. Keller and G. P. Carman, IEEE Transactions on Antennas and Propagation **63** (8), 3335-3344 (2015).
30. M. Zaeimbashi, H. Lin, C. Dong, X. Liang, M. Nasrollahpour, H. Chen, N. Sun, A. Matyushov, Y. He, X. Wang, C. Tu, Y. Wei, Y. Zhang, S. S. Cash, M. Onabajo, A. Shrivastava and N. Sun, IEEE Journal of Electromagnetics, RF and Microwaves in Medicine and Biology **3** (3), 206-215 (2019).
31. X. Liang, H. Chen, N. Sun, H. Lin and N. X. Sun, in *2018 IEEE International Symposium on Antennas and Propagation & USNC/URSI National Radio Science Meeting* (IEEE, 2018), pp. 2189-2190.
32. J. Xu, C. M. Leung, X. Zhuang, J. Li, S. Bhardwaj, J. Volakis and D. Viehland, Sensors **19** (4), 853 (2019).
33. D. Shuxiang, L. Jie-Fang and D. Viehland, IEEE Transactions on Ultrasonics, Ferroelectrics, and Frequency Control **50** (10), 1236-1239 (2003).
34. S. Dong, J. F. Li, D. Viehland, J. Cheng and L. E. Cross, Applied Physics Letters **85** (16), 3534-3536 (2004).
35. C. M. Leung, X. Zhuang, J. Xu, M. Gao, X. Tang, J. Li, P. Zhou, G. Srinivasan and D. Viehland, Journal of Physics D: Applied Physics **52** (6), 065003 (2018).
36. J. Zhai, J. Li, S. Dong, D. Viehland and M. I. Bichurin, Journal of Applied Physics **100** (12), 124509 (2006).
37. C. M. Leung, X. Zhuang, D. Friedrichs, J. Li, R. W. Erickson, V. Laletin, M. Popov, G. Srinivasan and D. Viehland, Applied Physics Letters **111** (12), 122904 (2017).
38. C. M. Leung, X. Zhuang, J. Xu, J. Li, G. Srinivasan and D. Viehland, Applied Physics Letters **110** (11), 112904 (2017).
39. J. Zhang, W. Zhu, D. Chen, H. Qu, P. Zhou, M. Popov, L. Jiang, L. Cao and G. Srinivasan, Journal of Magnetism and Magnetic Materials **473**, 131-135 (2019).
40. Y. Niu and H. Ren, Applied Physics Letters **118** (4), 044101 (2021).
41. M. Fiebig, Journal of Physics D: Applied Physics **38** (8), R123-R152 (2005).
42. C.-W. Nan, M. I. Bichurin, S. Dong, D. Viehland and G. Srinivasan, Journal of Applied Physics **103** (3) (2008).
43. G. Srinivasan, E. T. Rasmussen, J. Gallegos, R. Srinivasan, Y. I. Bokhan and V. M. Laletin, Physical Review B **64** (21), 214408 (2001).
44. M. B. Moffett, A. E. Clark, M. Wun-Fogle, J. Linberg, J. P. Teter and E. A. McLaughlin, The Journal of the Acoustical Society of America **89** (3), 1448-1455 (1991).
45. Y. Niu and H. Ren, Applied Physics Letters **118** (26), 264104 (2021).
46. A. E. Clark and H. S. Belson, Physical Preview B **5**, 3642 (1972).
47. J.-M. Hu, T. Nan, N. X. Sun and L.-Q. Chen, MRS Bulletin **40** (9), 728-735 (2015).
48. G.-T. Hwang, H. Palneedi, B. M. Jung, S. J. Kwon, M. Peddigari, Y. Min, J.-W. Kim, C.-W. Ahn, J.-J. Choi and B.-D. Hahn, ACS applied materials and interfaces **10** (38), 32323-32330 (2018).
49. D. Gibson, CREG Journal, 17-18 (2019).
50. L. J. Chu, Journal of Applied Physics **19** (12), 1163-1175 (1948).
51. G. Xu, S. Xiao, Y. Li and B. Wang, Physics Letters A **385**, 126959 (2021).